# Towards On-Chip Integrated Optical Quantum Frequency Combs

Lucia Caspani[1,2], Christian Reimer[1], Michael Kues[1], Piotr Roztocki[1], Matteo Clerici[1,3], Benjamin Wetzel[1,4], Yoann Jestin[1], Marcello Ferrera[1,2], Marco Peccianti[1,4], Alessia Pasquazi[1,4], Luca Razzari[1], Brent E. Little[5], Sai T. Chu[6], David J. Moss[7], and Roberto Morandotti[1,8]

[1]Institut National de la Recherche Scientifique - Énergie Matériaux et Télécommunications, Université du Québec, 1650 Boulevard Lionel-Boulet, Varennes, Québec, Canada J3X 1S2.
[2]Institute of Photonics and Quantum Sciences, Heriot-Watt University, Edinburgh EH14 4AS, UK.
[3]School of Engineering, University of Glasgow, Glasgow G12 8LT, UK.
[4]Department of Physics and Astronomy, University of Sussex, Falmer, Brighton BN1 9RH, UK.
[5]Xi'an Institute of Optics and Precision Mechanics of CAS, Xi'an 710119, China.
[6]Department of Physics and Material Science, City University of Hong Kong, Tat Chee Avenue, Hong Kong, China
[7]School of Electrical and Computer Engineering, RMIT University, Melbourne, Victoria 3001, Australia.
[8]Institute of Fundamental and Frontier Sciences, University of Electronic Science and Technology of China, Chengdu 610054, China.

**Abstract:** Recent development in quantum photonics allowed to start the process of bringing photonic-quantum-based systems out of the lab into real world applications. As an example, devices for the exchange of a cryptographic key secured by the law of quantum mechanics are currently commercially available. In order to further boost this process, the next step is to migrate the results achieved by means of bulky and expensive setups to miniaturized and affordable devices. Integrated quantum photonics is exactly addressing this issue. In this paper we briefly review the most recent advancements in the generation of quantum states of light (at the core of quantum cryptography and computing) on chip. In particular, we focus on optical microcavities, as they can offer a solution to the issue of low efficiency (low number of photons generated) typical of the materials mostly used in integrated platforms. In addition, we show that specifically designed microcavities can also offer further advantages, such as compatibility with existing telecom standard (thus allowing to exploit the existing fiber network) and quantum memories (necessary in turns to extend the communication distance), as well as longitudinal multimode character. This last property (i.e. the increased dimensionality necessary for describing the quantum state of a photon) is achieved thanks to the generating multiple photon pairs on a frequency comb corresponding to the microcavity resonances. Further achievements include the possibility to fully exploit the polarization degree of freedom also for integrated devices. These results pave the way to the generation of integrated quantum frequency combs, that in turn may find application as quantum computing platform.

**1. Introduction**:

Quantum photonics is playing an increasing role in the research community nowadays, and some applications already reached the commercial stage. Indeed, one of the preferred carriers for quantum information processing, and for communication in particular, is the photon. This is mainly due to the photon's low decoherence that allows to preserve its quantum state for long time, and thus for long propagation distances.

The main building blocks of quantum information, especially for computing and communication, are entangled photons (pairs of photons that share a single and not separable wave function) and single photons (nonclassical states of the electromagnetic field composed by one and only one photon). The most exploited method for generating these states is spontaneous parametric down conversion (SPDC) in second-order nonlinear crystals [1–5]. In this process a photon from an excitation field (the only input) is annihilated within the nonlinear crystal and two daughter twin photons (called signal and idler) are generated (Fig. 1A). One of the main advantage of SPDC is its versatility as it can be exploited for generating either entangled photons in different degrees of freedom (e.g. polarization, position, orbital angular momentum and more) or heralded single photons. In this last case, the property of the two photons of being generated exactly at the same time can be exploited for heralding the presence of one single photon on the idler beam upon detection of a photon in the signal beam.

While SPDC is nowadays a common tool in many quantum optical labs for generating quantum states of light, it typically relies on bulky setups and large efforts have been dedicated in the past years for reducing the footprint of these systems [6–8]. Following a trend typical for classical optics, the first step has been the generation of quantum states in optical fibers. Squeezing, quantum correlations, and entangled photon pair generation in different variables have all been achieved in fibers and we refer the reader to the thorough overview on this topic reported in [9]. More recently, with the development of integrated optics it became clear that the potential of quantum optics may fully be harnessed only in conjunction with the development of an appropriate integrated platform [10]. On the one hand photonic chips able to perform the logic ports at the basis of quantum computing have been demonstrated [11,12]. On the other, an intense research activity has been dedicated to implement on chip sources of single and entangled photons [13,6,14–16,7,17]. A detailed review on the genesis and evolution of integrated quantum optics can be found in [18].

For practical and ultimately widespread implementation, on-chip devices compatible with electronic integrated circuit technology offer great advantages in terms of low cost, small footprint, high performance and low energy consumption [19]. Furthermore, it is possible to manufacture a large number of such devices on a single chip, in a fully scalable platform. These devices can then be used individually or even be combined to ensembles. The output of the chip devices can then be fiber coupled to allow easy read-out and further manipulation, as well as compatibility with existing telecommunications infrastructures. However, materials used in silicon on-chip technology usually show only a third-order nonlinearity. In $\chi^{(3)}$ media photon pair generation is achieved through spontaneous four-wave mixing (SFWM), where two photons of the excitation field are annihilated for generating two daughter photons (Fig. 1B). While there is no fundamental restriction to use third-order nonlinear devices, they however suffer from other drawbacks such as lower nonlinear coefficients. As mentioned before, one solution was offered by optical fibers, where the low nonlinearity is compensated by the long interaction distance allowed by the mode confinement. Alternatively, a stronger interaction can be achieved exploiting resonant effects because of the field enhancement that they provide. In particular, triply resonant $\chi^{(3)}$ cavities provide a pair production rate improvement that scales as the sixth power of the cavity enhancement factor. To this purpose, slow-light photonic crystal waveguides [20] or microcavities [21] offer the possibility to combine the small footprint with a reasonable high generation rate. Furthermore, exploiting optical cavities for the generation of photon pairs allows to reduce the photon pair bandwidth [22,23], eventually matching the bandwidth required by quantum memories without the need of narrowband filters, which reduce

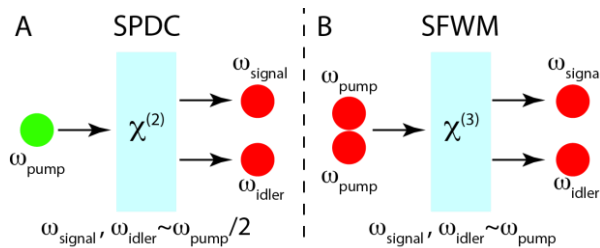

Figure 1. Spontaneous processes for the generation of (entangled) photon pairs in second-order (A) and third-order nonlinear media, where one or two photon, respectively, of the pump field are annihilated for generating two daughter photons.

the pair production rate.

The compatibility with quantum memories is fundamental in order to allow long-distance quantum key distribution (QKD). Indeed, the fiber losses limit the maximum distance achievable to few hundreds km. As mentioned before, standard amplifiers cannot be exploit to boost the propagation distance, since this will inevitably compromise the protocol security. A possible solution relies on quantum repeaters, i.e. repeater stations in which the signal is not amplified but rather the entanglement of a photon pair over a distance $L$ is achieved by proper combination of two entangled pairs, each of which extends over a distance $L/2$ [24]. However, a necessary requirement for quantum repeaters is the possibility to store the quantum state of the photon, and since quantum memories are typically based on atomic transitions that have linewidths on the order of 10 to 100 MHz [24], also the photon pairs must be featured by such bandwidths. Many of the sources based on integrated resonators have so far failed to achieve the narrow linewidths compatible with quantum memories because of their relatively modest Q-factors [14,16,25–28]. On the other hand, narrow linewidth sources can be achieved exploiting extremely high Q-factor cavities, however these are fundamentally incompatible with large-scale integration [6,15,29–31].

Another important feature allowed by the use of optical cavities for the generation of quantum states, is the possibility to achieve multimode entanglement. One of the most exploited degree of freedom for photon entanglement is polarization, a 2-dimensional variable, and therefore the quantum state of the photon is described by a 2-dimensional quantum systems or qubit (quantum bit). However, for variables with a dimensionality higher than 2 (multimode configuration) the photon state is described by a $D$-dimensional quantum system, thus passing from qubits to qu$D$its. The higher dimensionality allows for example to increase the quantity of information carried by a single photon. Different high dimensional variables, either discrete such as orbital angular momentum [32,33], or continuous such as time (or its conjugate, frequency) [34,35] and space (or its conjugate, transverse wave vector) [36] have been considered and investigated in the framework of bulk optics mainly with $\chi^{(2)}$ media.

However, almost all third-order devices are to date limited to the generation of two-mode single correlated states, and multiple correlated photon generation have only recently been investigated on chip [26,37,38]. In the next section, we report on our efforts towards the first realization of a wavelength-multiplexed source of heralded single photons on a chip compatible with both quantum memories and electronic chip fabrication processes (CMOS – complementary metal-oxide semiconductor – compatibility) [38].

## 2. Frequency comb of correlated photon pairs and single heralded photons: wavelength multiplexing

The requirements of a scalable and commercializable source for practical use include long-term operational stability and insensitivity to environmental changes, compatibility with quantum memories, operation at telecomm wavelengths (near 1550 nm), and the possibility of using large-scale electronic chip fabrication processes (CMOS). The development of such a source thus faces significant challenges, since on the one hand, quantum memories are typically associated with high-Q resonators, but external pumping of such high-Q microring resonators leads to meta-stable pump configurations due to thermal and environmental changes. While it is possible to lock an external excitation field to a single resonator, for example by using active electronic locking, not only does this dramatically increase the complexity of the system, rendering it incompatible with full integration on a chip, but it can actually lead to very unstable operation.

In order to solve this critical and difficult issue of locking the excitation field to ring resonances, we have developed a novel pumping scheme relying on passive optical feedback in a nested cavity-pumping configuration. By embedding the microring resonator in an external active cavity, it is directly implemented within the pump laser, where lasing is self-started by the amplified spontaneous emission of the active gain medium (fiber amplifier) employed and is then sustained by re-injecting a single ring resonance (the pump, selected by a tunable narrowband filter) into the amplifier (see Fig. 2).

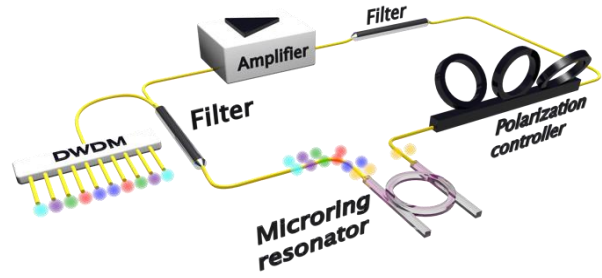

Figure 2. Experimental setup for the generation of multiplexed photon pairs in the self-locked pumping scheme: the pump photons are filtered and redirected to the amplifier for sustaining the lasing, while correlated photon pairs are demultiplexed by means of a dense wavelength division multiplexing (DWDM) filter before detection.

Excited by this stable pump configuration, a frequency comb of photons is generated through SFWM, where two $\nu_0$ excitation photons are annihilated and two photons at $\nu_0+n\Delta\nu$ and $\nu_0-n\Delta\nu$, respectively, are generated. Here $\nu_0$ is the pump frequency, $\Delta\nu$ is the microring free spectral range (FSR, 200 GHz), and $n$ is an integer. Since the signal/idler photons are generated in a single quantum process, their temporal correlation is expected to give a peak at zero delay. This peak is however broadened by the fact that once generated the photons can exit the cavity at different times, following the characteristic lifetime of the cavity determined by the ring resonator linewidth. In particular, for typical Lorentzian resonances it is thus expected a signal-idler temporal correlation of the form $C_{si}(\tau) \propto \exp(-2\pi\delta\nu|\tau|)$, where $\delta\nu$ is the cavity linewidth and $\tau$ is the signal-idler delay [22]. The correlation function is typically measured by detecting signal and idler photons with single photon detectors able to determine their arrival time with a resolution better than the cavity lifetime, and then determining the coincidence rate as a function of the signal-idler delay.

In order to achieve CW oscillation, preferable for generating high-coherence photon pairs, the FSR of the external cavity has to be large enough, and ideally larger than the microring linewidth ($\delta\nu$=140 MHz), so that only a single line of the external cavity oscillates. Note that a long pump coherence time corresponds to a large alphabet (the dimensionality is basically given by the ratio between the pump coherence time and the photon pair bandwidth, in turn determined by the cavity lifetime), i.e. high number of bits of information per photon [35].

When using an amplifier with a long gain medium (e.g. standard EDFA) the external cavity length is of the order of 30 meters, leading to an FSR well below the FWHM of the ring resonator. This leads to the oscillation of many lines (~24) of the external cavity, resulting in a chaotic lasing condition [39]. In addition, this chaotic pump behavior leads to the formation of isolated high peak-power pulses, which significantly lowers the threshold for optical parametric oscillation (OPO). Indeed, low-threshold OPO in this pump configuration was demonstrated at an average input power of only 6 mW [39]. In the unstable pump configuration, pulses oscillate inside the pump cavity with a repetition rate defined by the FSR of the external cavity (6 MHz), see Fig. 3A. These high-power peaks lead to the generation of multiple photon pairs per pulse (even at very low average pump powers), that can lead to a spread of the coincidence peak, when for example the detected signal and idler photons do not belong to a single pair, but are generated in different downconversion events separated in time. This in turn can be seen in the coincidence measurement, where this effect manifests in the superposition of a narrow and a wide peak, see Fig. 3B. The narrow peak corresponds to the real signal/idler coincidences, while the wide peak is caused by coincidences between uncorrelated photon pairs generated within the same pulse.

In order to enable CW oscillation even with a long external cavity, we have inserted a narrowband Fabry-Perot filter selecting only one external cavity line (see detailed experimental setup in Fig. 4). In this case a CW oscillation can be achieved, as shown in the measured radiofrequency (RF) spectrum (Fig. 5A). With a CW pump, only a narrow coincidence peak is visible as shown in Fig. 5B, underlining that the issue of multi-photon generation is eliminated in the stable CW pump configuration. It is worth noting that the issue of stability, i.e. the locking between the external pump laser and the ring resonances, is much more relevant for the high Q-factor microcavities needed to achieve high generation efficiencies and to match the narrow linewidths required for atomic-based quantum memories. For very high Q-factor ring resonators, such as the one employed in this work, the locking between the external CW laser and the resonance is usually lost after a few minutes, especially at the low pump powers typically used for photon pair generation. In our configuration instead, any thermal fluctuations are passively locked and the pump scheme allows for long-term stability, with no need for external, active feedback. One of the most promising features of the developed scheme is its potential to be fully monolithically integrated, enabling the practical implementation of microresonator-based quantum light sources.

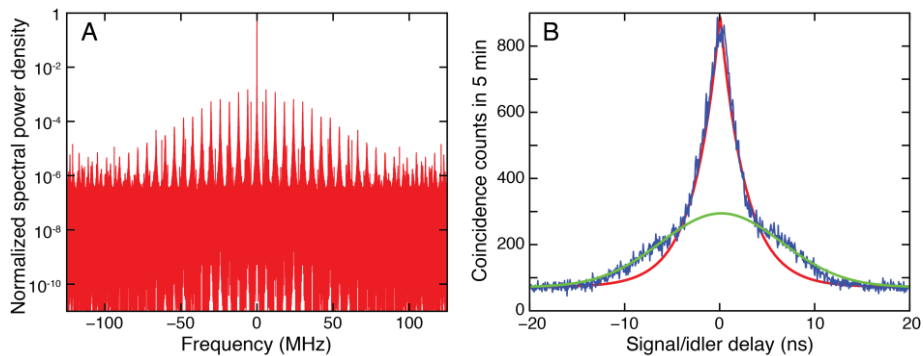

Figure 3. A. Radiofrequency spectrum of the unstable pump configuration, showing a clear pulsed operation with 6 MHz repetition rate. B. Coincidence peak measured in the unstable pump configuration (blue curve). A superposition of two peaks is visible, where the narrow signal/idler peak (fitted by the $C_{si}$ function, in red) is superimposed with a larger peak (fitted by a Gaussian function, in green).

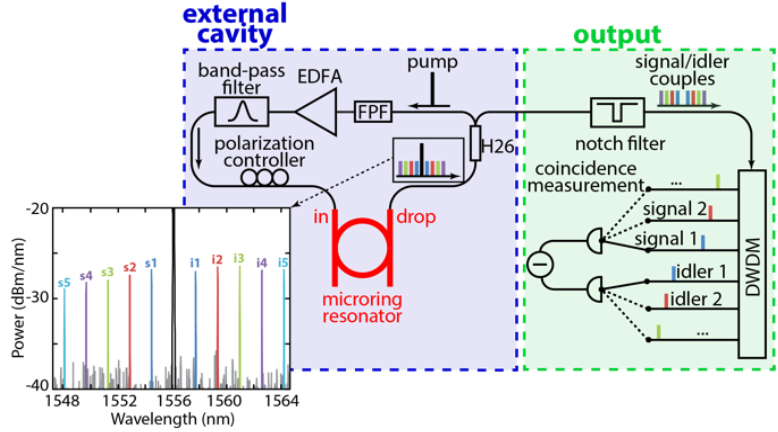

Figure 4. Detailed experimental setup showing the Fabry-Pérot filter (FPF) used to select a single external cavity resonance for achieving true CW pumping. The inset shows the spectrum of the generated photons (the black line is the pump). Adapted from Ref. [38].

By using our novel self-locked excitation scheme, we demonstrated an integrated photon pair source that generates multiple, simultaneous, and independent photon pairs multiplexed and distributed on a frequency comb grid that is compatible with the ITU channel spacing of optical fiber communications networks. In fact, although we limited our investigation to five channel pairs, our device is capable of generating photon pairs on a comb over a bandwidth that exceeds the telecom C and L bands (1530 nm to 1620 nm), corresponding to more than 80 different signal/idler channel pairs. The source is highly stable, operating continuously for several weeks with less than 5% fluctuation in photon flux, without any active stabilization thanks to the self-locked scheme. Using an integrated high-Q (1.375 million) microring resonator, we realized the generation of heralded single photons on five independent channels (centered at standard telecom channels, separated by 200 GHz). Measurements of the signal-idler correlation function for all resonance pair configurations of the first five resonances around the excitation frequency demonstrate the correlated photon pair emission only on resonance pairs spectrally symmetric to the pump frequency (Fig. 6). The source also features linewidths (110 MHz) that are orders of magnitude narrower than previous sources based on integrated ring resonators, and that are thus compatible with atomic-based quantum memories. Finally, the device is based on a platform that is compatible with electronic computer chip technology (CMOS). All together, these features potentially mark a substantial step forward to achieve stable, integrated and CMOS-compatible multi-mode sources for quantum optical applications. While in this particular case, only heralded single photons were generated, it has been shown that for example time-bin entanglement can be generated using SFWM on chip [27,28].

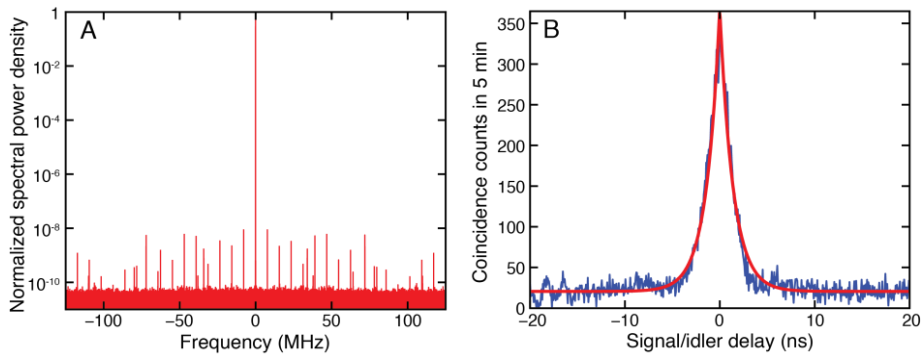

Figure 5. A. RF spectrum of the pump configuration with the long EDFA and narrowband FP filter showing clear CW operation. The noise at non-zero frequencies is caused by the limited dynamic range of the oscilloscope. D. Coincidence peak measured in the CW pump configuration using the long EDFA with the FP filter (blue curve), together with the corresponding $g^{(2)}$ fit (red curve).

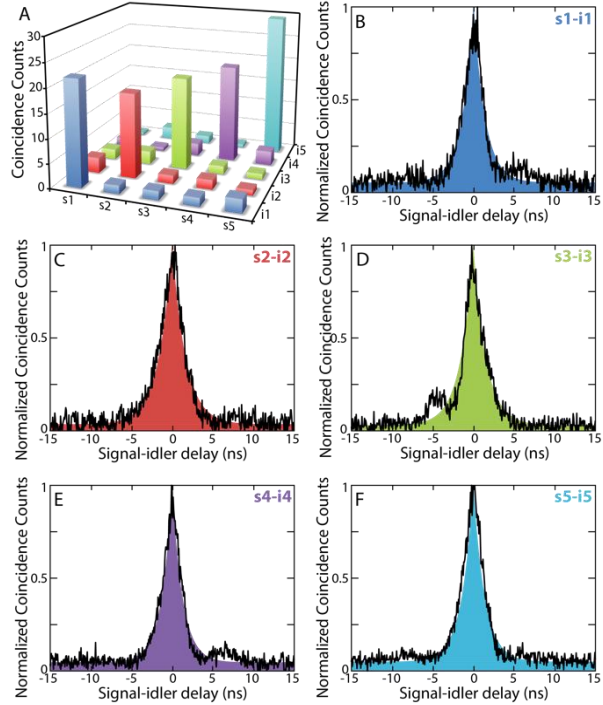

Figure 6. A. Coincidence count rate measured at all the signal/idler combinations. Significant coincidence counts (corresponding to a peak) are only visible between symmetric channels. B-F Normalized coincidence peaks (corrected for the dark counts due to the detector noise) measured for the 5 symmetric channel pairs (black curves) with the relative $C_{si}$ fit (solid-shaded curves).

## 3. Cross-polarized photon pair generation

As discussed in Section 2, entanglement generation using second-order nonlinear interactions has made use, in certain implementations, of polarization as an important degree of freedom to entangle photons between different modes. While most applications exploiting nonlinear processes in either bulk media or fiber-based devices have extensively relied on the electric field polarization as a fundamental degree of freedom to achieve novel nonlinear functionalities, in contrast, integrated third-order devices have not been able to exploit all of these degrees of freedom. Many on-chip key components have been realized, however, to date polarization has not been fully exploited as a degree of freedom for integrated nonlinear devices. In particular, frequency conversion based on orthogonally polarized beams has not yet been demonstrated on chip. Although multi-polarization processes, e.g. Type-II spontaneous FWM where the two excitation photons come from two respective polarization fields, could in principle be achieved on-chip by using two orthogonally-polarized pump fields, similarly to the fiber case [40,41], the overlapping and dominant stimulated processes generally make spontaneous FWM experimentally undetectable. Furthermore, to achieve efficient frequency conversion in a small footprint, integrated nonlinear processes are often enhanced through the use of cavities or photonic crystal waveguides, but these structures are typically designed for single polarization operation, as most integrated waveguides show a strong polarization-dependent dispersion and loss. Achieving efficient Type-II spontaneous FWM on a chip therefore requires structures that not only operate on two orthogonal polarizations with specific dispersion properties, but also provide nonlinear enhancement while suppressing competing stimulated processes.

In what follows, we report on our demonstration of frequency mixing between orthogonal polarization modes in a compact integrated microring resonator and the demonstration of a bi-chromatically pumped optical parametric oscillator [42]. Operating the device above and below threshold, we directly generate orthogonally polarized beams, as well as photon pairs, respectively, that can find applications, for example, in optical communication and quantum optics. Remarkably, by slightly adjusting the pump configuration presented in Section 2, the self-locked scheme also enables the stable excitation of two or more pump ring lines at the same time. By exciting two resonances of orthogonal polarization modes we introduced a new type of spontaneous four-wave mixing (FWM) to the toolbox of integrated photonics. In particular, we demonstrated the first realization of Type-II spontaneous FWM, in analogy with Type-II SPDC in second-order media. This scheme allows, for the first time, the direct generation of orthogonally-polarized photon pairs on a chip.

In order to demonstrate this novel configuration, we used a microring resonator that operated on the fundamental transverse electric (TE) and transverse magnetic (TM) modes, both having similar yet slightly different dispersions. Cavity enhancement is provided by the high Q-factors of the TE and TM resonances (235,000 and 470,000, respectively), which readily enable high parametric gain at low pump powers. In our scheme, the suppression of stimulated FWM between the two excitation fields was obtained by generating a frequency offset of 70 GHz between the TE and TM resonances, while keeping the free spectral ranges (FSRs) of both modes almost identical (200.39 and 200.51 GHz in this experiment), thus allowing Type-II spontaneous FWM to take place at targeted resonances without the presence of stimulated FWM. This offset is generated by the slightly different dispersions of the TE and TM modes, resulting in different effective resonator lengths and hence different resonant frequencies. Energy conservation dictates that the stimulated FWM bands have to be symmetric with respect to the two pump frequencies. Due to the frequency offset between the TE and TM resonances the spectral position of the stimulated FWM gain does not overlap with the ring resonances, thus suppressing this process inside the microring resonator (Fig. 7). At the same time, the TE and TM mode dispersions have to be kept similar so that the difference in FSR between the two modes (120 MHz) is smaller than the bandwidth of the resonances, in order to achieve energy conservation for Type-II spontaneous FWM processes. Furthermore, the mismatch between the FSRs with respect to the resonator bandwidth has to be minimized in order to achieve high efficiency in Type-II FWM. Finally, the required phase-matching condition can be obtained by operating in a slightly anomalous dispersion regime for both modes.

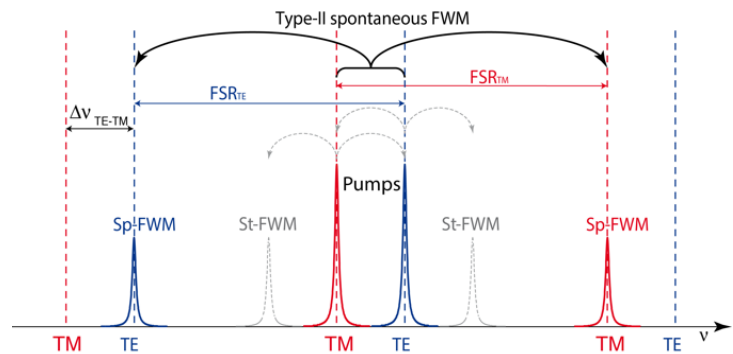

Figure 7. Scheme of the required relations between TE and TM FSRs in order to suppress stimulated FWM between the two cross-polarized pump fields. Figure from [42].

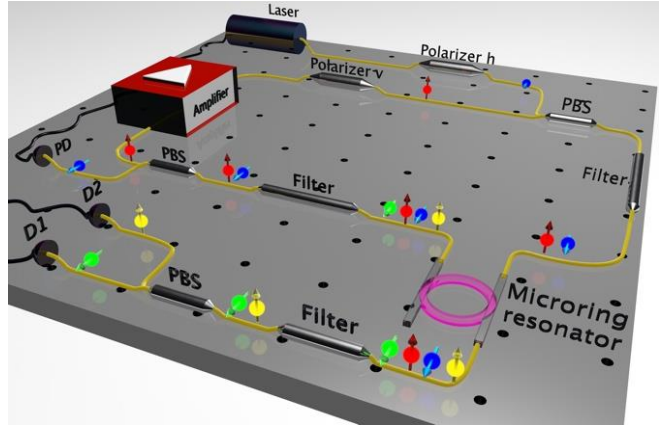

Figure 8. Experimental setup for achieving spontaneous four wave mixing between two cross-polarized pump fields. The scheme presented in previous section has been improved to allow pumping on two ring resonances. Figure from [42].

As discussed above, in order to simultaneously pump two resonances, we used a hybrid self-locked pumping approach, where the laser pumping the TE mode was directly built around the resonator, thus eliminating the need for active stabilization (Fig. 8). The microring resonator was embedded inside an external cavity that included a fiber amplifier and a wavelength filter. The amplified spontaneous emission of the fiber amplifier was transmitted through a band-pass filter (100 GHz) centered at the desired TE ring resonance and was then injected into the chip. Light coupled out of the drop port of the ring resonator was fed back to the amplifier, thereby closing the external pump cavity and promoting lasing on the TE mode. In order to allow self-locked lasing on only the TE polarization, while pumping the TM mode with an external laser (actively locked to the resonance using a feedback loop), polarizing beam couplers were placed before and after the ring resonator. This hybrid approach based on the use of one self-locked and one external excitation field permits pumping on both resonances in a very stable configuration and provides precise control over the individual pump powers.

When operated below the OPO threshold, our device directly generated orthogonally-polarized photon pairs. In order to characterize them and confirm the nature of the underlying nonlinear process, we performed photon coincidence measurements. The photon pairs generated in the TE and TM modes of the microring resonator were collected at the ring through port after appropriate filtering of both pump fields by means of a polarization-maintaining, high-isolation 200 GHz wide notch filter. The generated photons were then separated by a polarizing beam splitter and detected with single-photon detectors. We measured a clear coincidence peak with a coincidence-to-accidental ratio (CAR) of up to 12 without any background subtraction (Fig. 9A). As photon pairs can only be generated via spontaneous nonlinear processes, the measured photon coincidences give a strong indication that the photon pairs are generated through Type-II spontaneous FWM, with stimulated processes being successfully suppressed. The power-scaling behavior provides further insight into the process associated with the generation of the photon pairs. Only when one photon from each pump field is used to create two daughter photons, it becomes possible to directly generate orthogonally-polarized photon pairs. Therefore, the coincidence counts (CC) are expected to scale with the product of both pump powers (CC $\propto$ PTE x PTM). If the power of one pump field is kept constant, and the power of the second one is increased, a linear scaling behavior is predicted for Type-II spontaneous FWM, whereas if the power of both pump fields is simultaneously increased (with constant power ratio), a quadratic scaling is expected instead. No coincidences (within the noise) were measured when the ring was not pumped or pumped with the TE field alone, where the non-zero counts are due to the dark counts of the detector. A clear linear scaling behavior is visible (Fig. 9B) with increasing TM pump power and constant TE power, while a quadratic (without linear contribution) scaling is observed with increasing balanced pump powers. The presence of Raman scattering can be neglected, as the signal and idler frequencies do not overlap significantly with the Raman gain spectrum. This is also experimentally confirmed by the absence of any linear contribution to the power scaling that would arise from any Raman scattering (Fig. 9B).

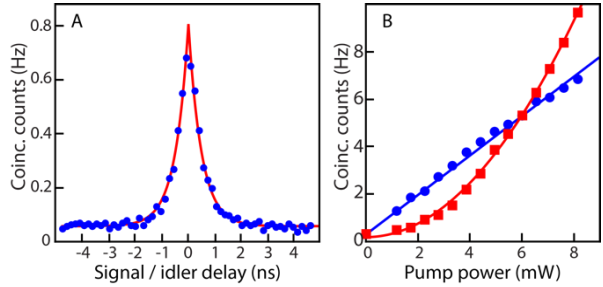

Figure 9. A. Coincidence peak of the cross-polarized photon pairs (blue points). The red curve is the $C_{si}$ fit. B. Coincidence count scaling as a function of the pump powers for balanced pumps (red squares, with quadratic fit – red curve) and when the power of one pump is kept constant at 6 mW (blue circles, with linear fit – blue curve). Figure adapted from [42].

From the coincidence measurement, shown in Fig. 9A, we extract a measured photon bandwidth of 320 MHz (black line), which is in good agreement with the resonator bandwidth of 410 MHz associated with the particular ring resonator used in this experiment, where the difference can be explained by the timing jitter of the detectors and electronics that results in a small temporal broadening of the measured peak. It is worth noting that the narrow linewidth, required for several quantum applications, is intrinsically achieved inside the resonator and cannot be directly realized in non-resonant waveguides or fiber-based architectures. The measured CAR of up to 12 is limited by loss, dark counts and the quantum efficiency of the detectors as well as by the photons generated through Type-0 SFWM of the individual pumps. These are issues that can be easily addressed in the future by optimizing the device dispersion. We measure a coincidence rate of around 4 Hz at 5 mW balanced pump power at the input of the chip (5mW is the highest achievable pump power featured by a CAR above 10). Considering all losses of the detection system (8.5 dB for both signal and idler) as well as the quantum efficiency of the detectors (5% and 10%, respectively), this corresponds to a pair production rate of 40 kHz and a pair production probability of $1.48 \times 10^{-12}$, accounting for the 1.6 dB coupling loss of the pump into the chip. By using better detectors and eventually implementing low-loss filtering on chip, the measured coincidence rate can be further increased to approach the production rate. Finally, the production of cross-polarized photon pairs is not limited to only the adjacent resonances, but the generation of frequency-multiplexed cross-polarized photon pairs is also possible. Indeed, we measured cross-polarized photon pairs over 12 resonance pairs, limited by the available filters, each with pair production rates above 20 kHz at 5 mW balanced pump power.

Having achieved Type-II spontaneous FWM in an integrated platform, we provide more access to polarization as a degree of freedom for integrated third-order spontaneous nonlinear interactions. Furthermore, since with our scheme two different FWM processes become accessible on the same chip, our device opens up the possibility of using, for example, Type-0 and Type-II FWM simultaneously to generate complex quantum optical states (for example, multi-entangled states) on a compact platform. All of these characteristics highlight the significant potential of our device for quantum optical applications.

## 4. Conclusions and Outlook
With integrated quantum photonics reaching a high level of maturity, novel systems able to generate more complex quantum states, with respect to the standard two-photon entangled qubits, can now be envisioned in an integrated platform. For example, entanglement between two identical integrated sources has been demonstrated, which opens the door to implement multiple source and entangle them to form a larger state [43]. Furthermore, we believe that the future of integrated quantum optics should take advantage of the decennial know-how developed in the framework of bulk quantum optics, and capitalize on the most recent developments in quantum optics, rather than reproducing its historical development in an integrated fashion. To this purpose, one of the path that is emerging as very promising, also in light of the foreseen and desirable application of quantum technologies out-of-the-lab, relies on multimode quantum states. While these have been widely investigated, they still represent a niche with respect to the more common two-dimensional states based on photon polarization. On the one hand, multimode states represent a valid solution for increasing the amount of information carried by a single photon, thus directly addressing the issue of the bit rate in the exchange of a quantum cryptographic key. On the other hand, they have been investigated as a possible platform for quantum computing. Indeed, a particular class of multimode quantum states, known as cluster states [44], in which each mode is entangled with more than one of the other modes, has been recently proposed as a novel approach to quantum computing, exploiting the so called one-way quantum computing [44–46]. In the standard circuit quantum computing model, one requires to implement evolution and control on each individual qubit [47]: the quantum algorithms are performed trough the different transformations (logic gates) applied to qubits. This in turn, strongly hindered the scalability of such systems. Instead, in the one-way quantum computing the complexity is shifted from the state control and manipulation to its generation: the different algorithms are realized simply trough different measurements. On this regard, recent developments in bulk optics [48–53], indicate that quantum frequency combs may represent a viable solution for one-way quantum computing. In this light, our recent results on the generation of wavelength multiplexed photon pairs in microring resonators demonstrate that this is a venue that should be investigated in integrated platforms. In addition, the possibility to add polarization as degree of freedom for microring-based parametric processes add a further control tool for achieving this goal.

**Acknowledgments:**
This work was supported by the Natural Sciences and Engineering Research Council of Canada (NSERC) through the Steacie and Discovery Grants Schemes, and by the Australian Research Council (ARC) Discovery Projects program. C.R. and P.R. acknowledge the support of an NSERC Vanier Canada Graduate Scholarship and NSERC Alexander Graham Bell Canada Graduate Scholarship-Master's (CGS-M), respectively. M.K. acknowledges the support from the "Fonds de recherche du Québec – Nature et technologies" (FRQNT) through the MELS fellowship program. We acknowledge the support from the People Programme (Marie Curie Actions) of the European Union's FP7 Programme: L.C. for THREEPLE under REA grant agreement n° [627478], B.W. for INCIPIT under REA grant agreement n° [625466], M.C. for KOHERENT under REA grant agreement n° [299522], M.F. for ATOMIC under REA grant agreement n° [329346], M.P. for THEIA under REA grant agreement n° [630833], and A.P. for CHRONOS under REA grant agreement n° [327627]. S.T.C. acknowledges the support from the CityU SRG-Fd program #7004189.

**References**

1. W. H. Louisell, A. Yariv, and A. E. Siegman, "Quantum Fluctuations and Noise in Parametric Processes. I.," Phys. Rev. **124**, 1646–1654 (1961).

2. D. N. Klyshko, "Coherent Photon Decay in a Nonlinear Medium," Pis'ma Zh. Eksp. Teor. Fiz. **6**, 490 (1967).

3. S. A. Akhmanov, V. V. Fadeev, R. V. Khokhlov, and O. N. Chunaev, "Quantum noise in parametric light amplifers," Pis'ma Zh. Eksp. Teor. Fiz. **6**, 575–578 (1967).

4. S. E. Harris, M. K. Oshman, and R. L. Byer, "Observation of Tunable Optical Parametric Fluorescence," Phys. Rev. Lett. **18**, 732–734 (1967).

5. D. Magde and H. Mahr, "Study in Ammonium Dihydrogen Phosphate of Spontaneous Parametric Interaction Tunable from 4400 to 16 000 Å," Phys. Rev. Lett. **18**, 905–907 (1967).

6. E. Pomarico, B. Sanguinetti, N. Gisin, R. Thew, H. Zbinden, G. Schreiber, A. Thomas, and W. Sohler, "Waveguide-based OPO source of entangled photon pairs," New J. Phys. **11**, 113042 (2009).

7. R. Horn, P. Abolghasem, B. J. Bijlani, D. Kang, A. S. Helmy, and G. Weihs, "Monolithic Source of Photon Pairs," Phys. Rev. Lett. **108**, 153605 (2012).

8. K.-H. Luo, H. Herrmann, S. Krapick, B. Brecht, R. Ricken, V. Quiring, H. Suche, W. Sohler, and C. Silberhorn, "Direct generation of genuine single-longitudinal-mode narrowband photon pairs," New J. Phys. **17**, 73039 (2015).

9. G. Agrawal, *Applications of Nonlinear Fiber Optics*, 2nd ed. (Academic Press, 2008).

10. J. O'Brien, B. Patton, M. Sasaki, and J. Vučković, "Focus on integrated quantum optics," New J. Phys. **15**, 035016 (2013).

11. A. Politi, M. J. Cryan, J. G. Rarity, S. Yu, and J. L. O'Brien, "Silica-on-silicon waveguide quantum circuits," Science (80-. ). **320**, 646–649 (2008).

12. A. Politi, J. C. F. Matthews, and J. L. O'Brien, "Shor's quantum factoring algorithm on a photonic chip," Science (80-. ). **325**, 1221 (2009).

13. H. Takesue, Y. Tokura, H. Fukuda, T. Tsuchizawa, T. Watanabe, K. Yamada, and S. Itabashi, "Entanglement generation using silicon wire waveguide," Appl. Phys. Lett. **91**, 201108 (2007).

14. S. Clemmen, K. Phan Huy, W. Bogaerts, R. G. Baets, P. Emplit, and S. Massar, "Continuous wave photon pair generation in silicon-on-insulator waveguides and ring resonators," Opt. Express **17**, 16558–70 (2009).

15. J. U. Fürst, D. V. Strekalov, D. Elser, A. Aiello, U. L. Andersen, C. Marquardt, and G. Leuchs, "Quantum Light from a Whispering-Gallery-Mode Disk Resonator," Phys. Rev. Lett. **106**, 113901 (2011).

16. S. Azzini, D. Grassani, M. J. Strain, M. Sorel, L. G. Helt, J. E. Sipe, M. Liscidini, M. Galli, and D. Bajoni, "Ultra-low power generation of twin photons in a compact silicon ring resonator," Opt. Express **20**, 23100–23107 (2012).

17. N. Matsuda, H. Le Jeannic, H. Fukuda, T. Tsuchizawa, W. J. Munro, K. Shimizu, K. Yamada, Y. Tokura, and H. Takesue, "A monolithically integrated polarization entangled photon pair source on a silicon chip," Sci. Rep. **2**, 817 (2012).

18. S. Tanzilli, A. Martin, F. Kaiser, M. P. De Micheli, O. Alibart, and D. B. Ostrowsky, "On the genesis and evolution of Integrated Quantum Optics," Laser Photon. Rev. **6**, 115–143 (2012).

19. D. J. Moss, R. Morandotti, A. L. Gaeta, and M. Lipson, "New CMOS-compatible platforms based on silicon nitride and Hydex for nonlinear optics," Nature Phot. **7**, 597–607 (2013).

20. M. J. Collins, M. J. Steel, T. F. Krauss, B. J. Eggleton, and A. S. Clark, "Photonic Crystal Waveguide Sources of Photons for Quantum Communication Applications," IEEE J. Sel. Top. Quantum Electron. **21**, 205–214 (2015).

21. K. J. Vahala, "Optical microcavities.," Nature **424**, 839–46 (2003).

22. Z. Ou and Y. Lu, "Cavity Enhanced Spontaneous Parametric Down-Conversion for the Prolongation of Correlation Time between Conjugate Photons," Phys. Rev. Lett. **83**, 2556–2559 (1999).

23. K. Garay-Palmett, Y. Jeronimo-Moreno, and a B. U'Ren, "Theory of cavity-enhanced spontaneous four wave mixing," Laser Phys. **23**, 015201 (2013).

24. N. Sangouard, C. Simon, H. de Riedmatten, and N. Gisin, "Quantum repeaters based on atomic ensembles and linear optics," Rev. Mod. Phys. **83**, 33–80 (2011).

25. E. Engin, D. Bonneau, C. M. Natarajan, A. S. Clark, M. G. Tanner, R. H. Hadfield, S. N. Dorenbos, V. Zwiller, K. Ohira, N. Suzuki, H. Yoshida, N. Iizuka, M. Ezaki, J. L. O'Brien, and M. G. Thompson, "Photon pair generation in a silicon micro-ring resonator with reverse bias enhancement," Opt. Express **21**, 27826–27834 (2013).

26. R. Kumar, J. R. Ong, J. Recchio, K. Srinivasan, and S. Mookherjea, "Spectrally multiplexed and tunable-wavelength photon pairs at 1.55 μm from a silicon coupled-resonator optical waveguide," Opt. Lett. **38**, 2969–2971 (2013).

27. D. Grassani, S. Azzini, M. Liscidini, M. Galli, M. J. Strain, M. Sorel, J. E. Sipe, and D. Bajoni, "Micrometer-scale integrated silicon source of time-energy entangled photons," Optica **2**, 88 (2015).

28. C. Xiong, X. Zhang, A. Mahendra, J. He, D.-Y. Choi, C. J. Chae, D. Marpaung, A. Leinse, R. G. Heideman, M. Hoekman, C. G. H. Roeloffzen, R. M. Oldenbeuving, P. W. L. van Dijk, C. Taddei, P. H. W. Leong, and B. J. Eggleton, "Compact and reconfigurable silicon nitride time-bin entanglement circuit," Optica **2**, 724 (2015).

29. M. Förtsch, J. U. Fürst, C. Wittmann, D. Strekalov, A. Aiello, M. V Chekhova, C. Silberhorn, G. Leuchs, and C. Marquardt, "A versatile source of single photons for quantum information processing," Nat. Commun. **4**, 1818 (2013).

30. C.-S. Chuu, G. Y. Yin, and S. E. Harris, "A miniature ultrabright source of temporally long, narrowband biphotons," Appl. Phys. Lett. **101**, 051108 (2012).

31. F. Monteiro, a. Martin, B. Sanguinetti, H. Zbinden, and R. T. Thew, "Narrowband photon pair source for quantum networks," Opt. Express **22**, 4371 (2014).

32. J. Leach, B. Jack, J. Romero, A. K. Jha, A. M. Yao, S. Franke-Arnold, D. G. Ireland, R. W. Boyd, S. M. Barnett, and M. J. Padgett, "Quantum correlations in optical angle-orbital angular momentum variables," Science (80-. ). **329**, 662–665 (2010).

33. A. C. Dada, J. Leach, G. S. Buller, M. J. Padgett, and E. Andersson, "Experimental high-dimensional two-photon entanglement and violations of generalized Bell inequalities," Nature Phys. **7**, 677–680 (2011).

34. I. Ali Khan and J. Howell, "Experimental demonstration of high two-photon time-energy entanglement," Phys. Rev. A **73**, 031801 (2006).

35. I. Ali-Khan, C. Broadbent, and J. Howell, "Large-alphabet quantum key distribution using energy-time entangled bipartite states," Phys. Rev. Lett. **98**, 060503 (2007).

36. M. Kolobov, "The spatial behavior of nonclassical light," Rev. Mod. Phys. **71**, 1539–1589 (1999).

37. W. C. Jiang, X. Lu, J. Zhang, O. Painter, and Q. Lin, "Silicon-chip source of bright photon pairs," Opt. Express **23**, 20884 (2015).

38. C. Reimer, L. Caspani, M. Clerici, M. Ferrera, M. Kues, M. Peccianti, A. Pasquazi, L. Razzari, B. E. Little, S. T. Chu, D. J. Moss, and R. Morandotti, "Integrated frequency comb source of heralded single photons.," Opt. Express **22**, 6535–6546 (2014).

39. A. Pasquazi, L. Caspani, M. Peccianti, M. Clerici, M. Ferrera, L. Razzari, D. Duchesne, B. E. Little, S. T. Chu, D. J. Moss, and R. Morandotti, "Self-locked optical parametric oscillation in a CMOS compatible microring resonator: a route to robust optical frequency comb generation on a chip," Opt. Express **21**, 13333 (2013).

40. Q. Lin, F. Yaman, and G. Agrawal, "Photon-pair generation in optical fibers through four-wave mixing: Role of Raman scattering and pump polarization," Phys. Rev. A **75**, 023803 (2007).

41. E. Brainis, "Four-photon scattering in birefringent fibers," Phys. Rev. A **79**, 023840 (2009).

42. C. Reimer, M. Kues, L. Caspani, B. Wetzel, P. Roztocki, M. Clerici, Y. Jestin, M. Ferrera, M. Peccianti, A. Pasquazi, B. E. Little, S. T. Chu, D. J. Moss, and R. Morandotti, "Cross-polarized photon-pair generation and bi-chromatically pumped optical parametric oscillation on a chip," Nat. Commun. **6**, 8236 (2015).

43. J. W. Silverstone, R. Santagati, D. Bonneau, M. J. Strain, M. Sorel, J. L. O'Brien, and M. G. Thompson, "Qubit entanglement between ring-resonator photon-pair sources on a silicon chip," Nat. Commun. **6**, 7948 (2015).

44. R. Raussendorf and H. J. Briegel, "A One-Way Quantum Computer," Phys. Rev. Lett. **86**, 5188–5191 (2001).

45. N. C. Menicucci, P. van Loock, M. Gu, C. Weedbrook, T. C. Ralph, and M. a. Nielsen, "Universal Quantum Computation with Continuous-Variable Cluster States," Phys. Rev. Lett. **97**, 110501 (2006).

46. J. Zhang and S. L. Braunstein, "Continuous-variable Gaussian analog of cluster states," Phys. Rev. A **73**, 032318 (2006).

47. M. A. Nielsen and I. L. Chuang, *Quantum Computation and Quantum Information* (Cambridge University Press, 2000).

48. S. Yokoyama, R. Ukai, S. C. Armstrong, C. Sornphiphatphong, T. Kaji, S. Suzuki, J. Yoshikawa, H. Yonezawa, N. C. Menicucci, and A. Furusawa, "Ultra-large-scale continuous-variable cluster states multiplexed in the time domain," Nature Phot. **7**, 982–986 (2013).

49. O. Pfister, S. Feng, G. Jennings, R. Pooser, and D. Xie, "Multipartite continuous-variable

entanglement from concurrent nonlinearities," Phys. Rev. A **70**, 020302 (2004).

50. O. Pinel, P. Jian, R. M. de Araújo, J. Feng, B. Chalopin, C. Fabre, and N. Treps, "Generation and Characterization of Multimode Quantum Frequency Combs," Phys. Rev. Lett. **108**, 083601 (2012).

51. J. Roslund, R. M. de Araújo, S. Jiang, C. Fabre, and N. Treps, "Wavelength-multiplexed quantum networks with ultrafast frequency combs," Nature Phot. **8**, 109–112 (2013).

52. M. Pysher, Y. Miwa, R. Shahrokhshahi, R. Bloomer, and O. Pfister, "Parallel Generation of Quadripartite Cluster Entanglement in the Optical Frequency Comb," Phys. Rev. Lett. **107**, 030505 (2011).

53. M. Chen, N. C. Menicucci, and O. Pfister, "Experimental Realization of Multipartite Entanglement of 60 Modes of a Quantum Optical Frequency Comb," Phys. Rev. Lett. **112**, 120505 (2014).